# Machine learning identification of fractional-order vortex beam diffraction process*

GUO Yan, LYU Heng, DING Chunling, YUAN Chenzhi, JIN Ruibo
Hubei Key Laboratory of Optical Information and Pattern Recognition, Wuhan Institute of Technology, Wuhan 430205, China

**Abstract**
Fractional-order vortex beams possess fractional orbital angular momentum (FOAM) modes, which theoretically have the potential to increase transmission capacity infinitely. Therefore, they have significant application prospects in the fields of measurement, optical communication and micro-particle manipulation. However, when fractional-order vortex beams propagate in free space, the discontinuity of the helical phase makes them susceptible to diffraction in practical applications, thereby affecting the accuracy of OAM mode recognition and severely limiting the use of FOAM-based optical communication. Achieving machine learning recognition of fractional-order vortex beams under diffraction conditions is currently an urgent and unreported issue. Based on ResNet, a deep learning (DL) method of accurately recognizing the propagation distance and topological charge of fractional-order vortex beam diffraction process is proposed in this work. Utilizing both experimentally measured and numerically simulated intensity distributions, a dataset of vortex beam diffraction intensity patterns in atmospheric turbulence environments is created. An improved 101-layer ResNet structure based on transfer learning is employed to achieve accurate and efficient recognition of the FOAM model at different propagation distances. Experimental results show that the proposed method can accurately recognize FOAM modes with a propagation distance of 100 cm, a spacing of 5 cm, and a mode spacing of 0.1 under turbulent conditions, with an accuracy of 99.69%. This method considers the effect of atmospheric turbulence during spatial transmission, allowing the recognition scheme to achieve high accuracy even in special environments. It has the ability to distinguish ultra-fine FOAM modes and propagation distances, which cannot be achieved by traditional methods. This technology can be applied to multidimensional encoding and sensing measurements based on FOAM beam.

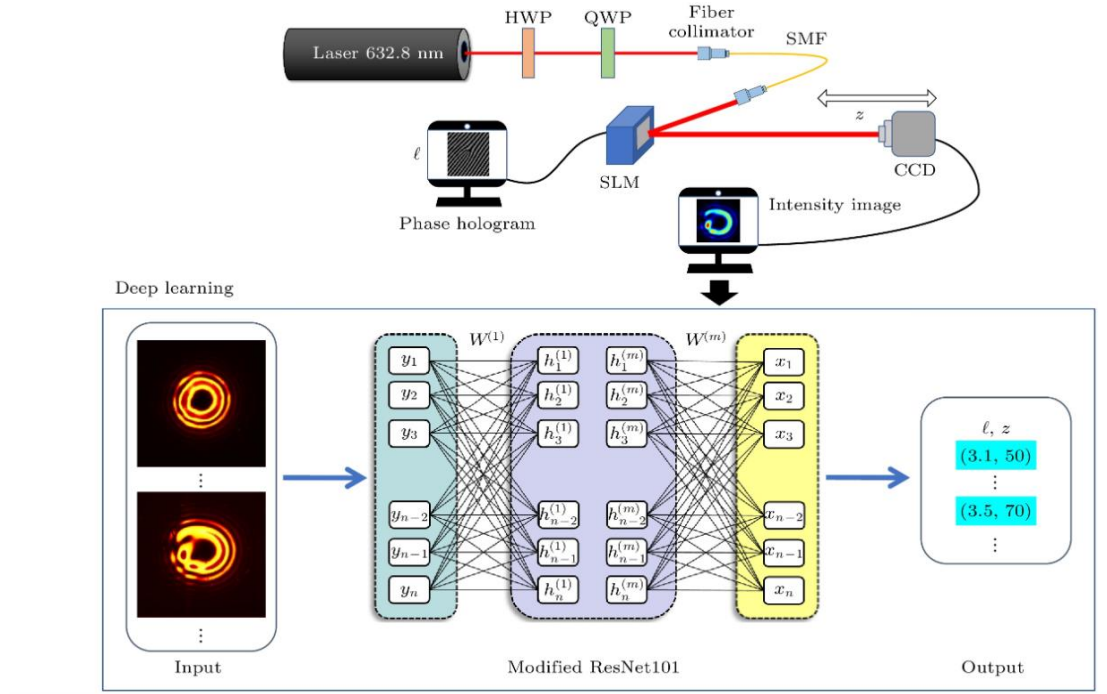

# 1. Introduction

Optical vortex is a special light field with spiral wavefront structure and definite orbital angular momentum (OAM) of photon[1–5]. In the 1970s, scientists proposed the concept of optical vortex, and did a lot of research on the annular intensity distribution characteristics of the vortex optical field and the phase singularity at the optical axis[6,7]. However, it was not until 1992 that Allen et al.[8] theoretically clarified the physical picture of the optical field of a vortex carrying orbital angular momentum, that is, a beam with a spiral phase wavefront and zero intensity at the center of the vortex. The vortex beam carries a phase factor of exp (i$\ell\theta$), where each photon carries $\ell\hbar$ orbital angular momentum, $\theta$ is the spatial azimuth, and $\ell$ is the topological charges (TC), so the vortex beam is also called orbital angular momentum beam[9]. Unlike the spin angular momentum, the orbital angular momentum can take any number, so the orbital angular momentum of a photon can be used to construct a high-dimensional Hilbert space. This feature makes it an ideal carrier of high-dimensional classical and quantum information. When the orbital angular momentum is a fraction, the beam becomes a fractional-order vortex beam with fractional orbital angular momentum. FOAM has unique physical properties, such as arbitrary radial notch, rich phase structure and higher modulation dimension[11]. These physical properties make the fractional vortex light have more control parameters, so it can carry more information, and has stronger coding ability and parameter control ability. Therefore, fractional vortex beams have been widely used in optical particle manipulation[12], optical information transmission[13,14], optical imaging[15], etc. In the application of fractional vortex beams, the identification of their OAM order is a core task. Traditional recognition methods include Mach-Zehnder interferometry[16,17], mode conversion[18,19] and machine learning[20–24]. Among them, machine learning method has unique advantages, which can take various external factors into account in the learning process and automatically recognize patterns, which is very helpful for the study of some complex physical phenomena[25–26]. However, the above study mainly deals with the identification of ordinary fractional vortex light, and does not involve the influence of diffraction. In fact, the influence of diffraction on the propagation of vortex beams is very obvious, especially for fractional-order vortex beams, the discontinuity of the spiral phase makes it easier to have strong diffraction in practical applications under the conditions of long-distance propagation and atmospheric turbulence, thus affecting the accuracy of OAM order identification[23,27]. How to realize the machine learning recognition of fractional order vortices under the condition of diffraction is still an urgent problem to be solved but rarely reported.

In this paper, the OAM mode is identified from the fractional topological charge $\ell$ and the propagation distance $z$ for the diffracted fractional vortex light. In the atmospheric turbulence environment, a 101-layer structure based on residual network (ResNet), including 100 convolutional layers and a fully connected layer, is used to detect the distorted fractional-order OAM mode. Experimental results show that the proposed method can accurately identify FOAM modes with a propagation distance of 100 cm, an interval of 5 cm, and a mode spacing of 0.1, with an accuracy of 99. 69%. In addition, the method has good generalization ability and can effectively resist interference in complex transmission environment, which provides a new idea for the application of FOAM beams in multi-dimensional coding and sensing measurement.

## 2. Generation of Fractional Vortex Beams under Diffraction Conditions

In this section, we first introduce the theory of vortex light generation, and then introduce the experiment of vortex light generation.

### 2.1 Theory of generation of fractional-order vortex beams.

When a Gaussian beam $\exp(-r^2/\omega_0^2)$ is incident on a spatial light modulator (SLM) and a phase mask $\exp(-i\ell\theta)$ is loaded on the SLM, the beam amplitude on the plane of SLM can be expressed as [28]

$$E_1(r, \theta) = \exp\left(-\frac{r^2}{\omega_0^2}\right) \exp(-i\ell\theta), \tag{1}$$

Where the topological charge number $\ell$ is a fractional value, $\omega_0$ is the Gaussian beam waist, $r$ and $\theta$ are the radial and azimuthal coordinates, respectively. For a fractional Gaussian vortex beam expressed by (1), we usually decompose the fractional vortex phase term into a basis of integer vortex phase terms:

$$\exp(-i\ell\theta) = \frac{1}{\pi}\exp(i\pi\ell)\sin(\pi\ell)\sum_{n=-\infty}^{+\infty}\frac{\exp(i\theta n)}{\ell - n}. \tag{2}$$

In the framework of paraxial approximation, the field distribution[29] of $E_1(r,\theta)$ after propagation can be calculated by using Collins integral equation:

$$E_2(r_1, \theta_1, z) = \frac{i}{\lambda B}\exp(-ikz)\int_0^{2\pi}\int_0^{\infty} E_1(r, \theta)$$
$$\times \exp\left\{-\frac{ik}{2B}[Ar^2 - 2rr_1\cos(\theta_1 - \theta) + Dr_1^2]\right\}rdrd\theta, \tag{3}$$

Where $r_1$ and $\theta_1$ are the radial and azimuthal coordinates in the output plane, $z$ is the propagation distance, $k = 2\pi/\lambda$ is the wave number, and $\lambda$ is the wavelength. The *ABCD* transfer matrix for light propagating in free space at distance $z$ is

$$\begin{pmatrix} A & B \\ C & D \end{pmatrix} = \begin{pmatrix} 1 & z \\ 0 & 1 \end{pmatrix}. \tag{4}$$

Substituting (1) and (2) into (3), the beam amplitude can be obtained:

$$
\begin{aligned}
E_2&(r_1, \theta_1, z) \\
&= \frac{1}{\lambda z} \exp(-ikz) \exp\left(-\frac{ikr_1^2}{2z}\right) \exp(i\pi\ell) \sin(\pi\ell) \\
&\times \sum_{n=-\infty}^{+\infty} \frac{i^{n+1} \exp(in\theta_1)}{\ell - n} \frac{b_1^n}{\varepsilon_1^{1+\frac{n}{2}}} \frac{\Gamma(n/2+1)}{\Gamma(n+1)} \\
&\times {}_1F_1\left(\frac{n+2}{2}, n+1, -\frac{b_1^2}{\varepsilon_1}\right).
\end{aligned}
\tag{5}
$$

Equation (5) represents the hypergeometric Gaussian mode, ${}_1F_1(\alpha,\beta,z)$ is a confluent hypergeometric function, $\Gamma(n)$ is the Gamma function, and $b_1$ and $\varepsilon_1$ are defined as

$$
b_1 = \frac{kr_1}{2z}, \quad \varepsilon_1 = \frac{1}{\omega_0^2} + \frac{ik}{2z}
\tag{6}
$$

Based on the above calculation, the transverse intensity distributions of vortex beam with different values of $\ell$ after propagating for different distances $z$ can be obtained. In actual communication, the spiral phase structure of vortex beams is easily distorted due to the existence of atmospheric turbulence, which leads to mode dispersion and intensity distribution distortion. Therefore, in this experiment, the Kolmogorov model with Von Karman turbulence spectrum is used to simulate the situation of spatial light modulator affected by atmospheric turbulence, thus realizing a distorted communication mode[30,31], whose distortion degree can be quantified by Fried parameter. The expression for the turbulent phase mask added on the SLM is[32,33]

$$
\Phi(x, y) = \mathbb{R}\left\{\mathcal{F}^{-1}\left(\mathbb{M}_{NN} \sqrt{\phi_{NN}(\kappa)}\right)\right\}
\tag{7}
$$

Where $\phi_{NN}(\kappa) = 0.023 r_0^{-5/3} \left(\kappa^2 + \kappa_0^2\right)^{-11/6} e^{-\kappa^2/\kappa_m^2}$ and Fried parameter $r_0 = \left(0.423 k^2 C_n^2 z\right)^{-3/5}$. $\mathbb{R}$ represents the real part of the complex field, and $\mathcal{F}^{-1}$ represents the inverse Fourier transform operation. In addition, $\kappa$, $\kappa_0$, and $\mathbb{M}_{NN}$ represent the spatial frequency, the central spatial frequency, and the encoded random matrix, respectively. $C_n^2$ is the atmospheric refractive index structure constant, and its value is used to express the turbulence intensity.

After loading both fractional-order vortex phase and turbulent phase masks on the SLM, the amplitude of the beam at the SLM plane becomes

$$E'_1(r,\theta) = E_1(r,\theta)\exp(i\varPhi(x,y)). \tag{8}$$

By substituting equation (8) into equation (5), the light field distribution $E'_2(r_1,\theta_1,z)$. after turbulence distortion can be obtained.

According to equation (8), the vortex beam can be simulated to obtain the vortex beams distribution under different $\ell$ and $z$ values. Fig. 2 shows the spatial distribution of the vortex beams without turbulence and with turbulence.

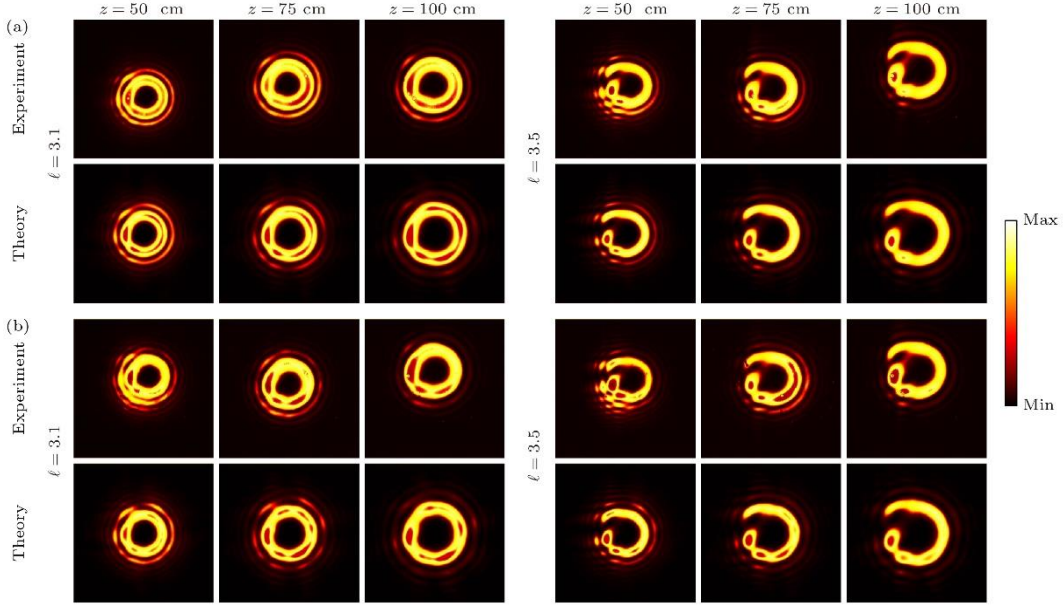

**Figure 2.** Spatial profiles of vortex beams with different topological charges $\ell$ and different propagation distances $z$: (a) Spatial distribution of distortionless modes without turbulence; (b) the spatial distribution of distortion modes affected by atmospheric turbulence. The first and third rows are the images acquired from the experiment, and the second and fourth rows represent the theoretically simulate.

2.2 Experimental generation of fractional-order vortex beams.

The experimental setup for the preparation of fractional order vortex beams under diffractive conditions is shown in Fig. 1. Firstly, the He-Ne laser beam with the wavelength of 632.8 nm is polarized by a half-wave plate (HWP) and a quarter-wave plate (QWP), and then coupled into a single-mode fiber to change its spatial mode into a pure Gaussian distribution. Then, an objective lens with a magnification of 10 × and an effective focal length of 17 mm is used to collimate the beam, and the collimated beam waist is around 2 mm. The collimated Gaussian beam is incident on the SLM and converted into a vortex beam. A computer-generated phase hologram is loaded on the SLM, and a turbulent phase is added to the hologram to simulate the turbulence in the atmosphere. Finally, the intensity image of the vortex beam is collected by a CCD camera and sent to a computer for training. The propagation distance is controlled by changing the position of the CCD. The range of $\ell$ is 1.0-9.9, $\Delta\ell$=0.1, the range of propagation distance $z$ is 50-100 cm, the step is 5 cm, and the turbulence intensity coefficient $C_n^2$ is $5\times10^{-10}$ mm$^{-2/3}$.

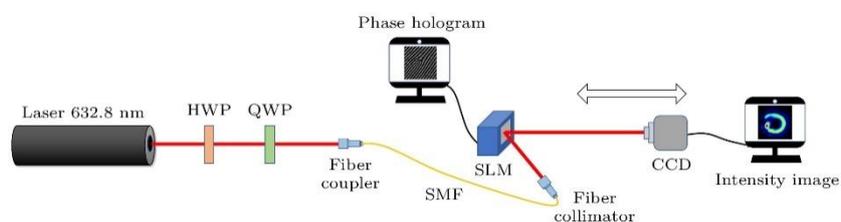

**Figure 1.** Diagram of experimental setup.

The fractional vortex beam not only shows phase singularity, but also has a dislocation in the radial direction, which will lead to the destruction of the central symmetry of the diffraction pattern, thus producing a series of effects. The spatial distribution of different topological charge numbers $\ell$ and different propagation distances $z$ is shown in Fig. 2, where Fig. 2(a) is the spatial distribution of the undistorted mode without turbulence, and Fig. 2(b) is the spatial distribution of the distorted mode under the influence of atmospheric turbulence. The first and third rows are the images acquired from the experiment, and the second and fourth rows represent the theoretically simulate. By comparing the experimental results with the theoretical images, it is found that the spatial profiles between them are very similar, which verifies the correctness of our theoretical model. It can be seen from Fig. 2 that the size of the central aperture and the diffraction aperture increases with the increase of the topological charge number of the vortex beam and the propagation distance. When the topological charge number is close to a half-integer, the radial gap is more obvious, and the light intensity on both sides of the radial gap is significantly greater than that at other positions.

**3. Identification of fractional-order vortex beams.**

3.1 Deep Learning Algorithm Design

In this paper, the ResNet transfer learning method is used, and the network structure is shown in Fig. 3. First, the previous network model parameters are used as the initialization values of the vortex light image training network model. ResNet-101 starts with a common convolutional layer, which is responsible for feature extraction of the input image, usually using a 3 × 3 convolutional kernel. After the convolution layer, the result of each convolution operation is normalized on each mini-batch by a batch normalization (BN) layer. Then, according to the characteristics of fractional vortex light image recognition, the rectified linear unit (ReLU) is used as the activation function, and the nonlinear transformation is introduced to increase the expressive power of the model, accelerate the training and convergence speed of the network, inhibit over-fitting, and improve the robustness of the model. The ReLU function has the advantages of simple calculation, no gradient saturation, and unilateral suppression, which can better extract image features. The model uses five residual modules to achieve deeper training and extract higher-level features. Each residual block is composed of multiple convolutional layers, batch normalization layers and activation function layers. Using skip connection and residual learning, the residual block can solve the problems of gradient disappearance and gradient explosion in deep networks, thus making the network easier to train and optimize. After the last residual block, the model adds a global Max pooling layer to compress the feature map into a vector, which reduces the number of parameters and improves the calculation speed. After that, a fully connected network is added, which consists of two fully connected layers, a random dropout layer is added before each fully connected layer, some parameters are randomly deleted to minimize overfitting, and the last layer of the pre-trained network is fully connected. Finally, the output of this layer is set to the classification number 99 of the fractional vortex light image data set in this paper, and the training accuracy is obtained by identifying the training set. The transfer learning toolkit (TLT)[34] is used in the whole algorithm, which uses the pre-trained model, so that we can achieve accurate recognition results with only fewer datasets, which is very efficient[22,35].

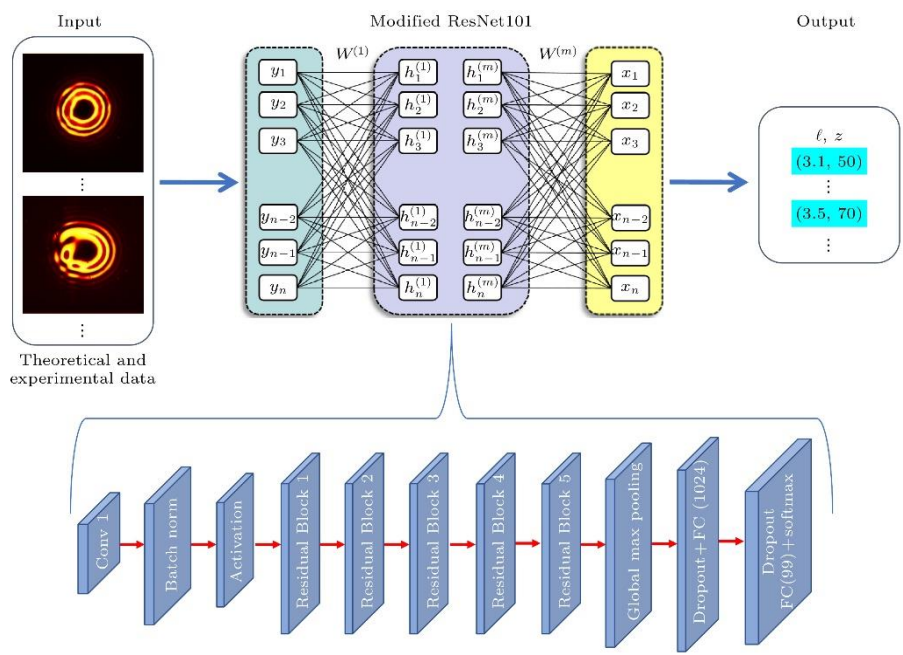

**Figure 3.** The deep learning algorithm. The deep learning network consists of the unaltered ResNet-101 bottom layer and our redesigned top layer.

According to the characteristics of the target task of multi-class classification in transfer learning, the categorical cross entropy loss function[36] is selected to evaluate the training results of each epoch:

$$\text{Loss} = -\sum_{i=1}^{n}(\hat{y}_{i1}\ln y_{i1} + \hat{y}_{i2}\ln y_{i2} + \cdots + \hat{y}_{im}\ln y_{im}), \qquad (9)$$

Where *n* is the number of samples, *m* is the number of classifications, $\hat{y}_{im}$ represents the true label (value 0 or 1), and $y_{im}$ is the predicted value of the *m*th given by the neural network. In the training process, the adaptive moment estimation (Adam) optimizer[37] is used to update the weights and bias parameters to minimize the loss function. The categorical cross entropy loss function can make full use of the label information to model the prediction probability of each class, maximize the prediction probability of the correct class, and impose a large penalty on the misclassified samples, thus prompting the model to pay more attention to the samples that are difficult to classify. The cross entropy loss function has strong generalization ability, and can predict well even on unknown data sets. In addition, it has good convex optimization, that is, when the probability value predicted by the model is close to the true label value, the value of the loss function will be smaller and smaller, and when the probability value predicted by the model is far away from the true label value, the value of the loss function will be larger and larger, thus contributing to the convergence of model training.

3.2 Fractional vortex beam recognition based on deep learning algorithm

After the machine learning model is constructed, the experimental intensity distribution images and the theoretical simulated intensity distribution images obtained in the 2.2 section are trained, wherein the experimental intensity distribution images and the theoretical simulated intensity distribution images constitute the whole data set according to 7:3, the training set contain 7920 images, the validation set contain

1980 images, the test set contain 990 images, and each image has a resolution of 480 × 360 pixels. The experiment is conducted on a computer equipped with an Intel® Core™ i5-7300HQ @2.5 GHz processor and an NVIDIA GeForce GTX 1050 Ti GPU (4 GB VRAM). The software environment consists of Python 3.9 and TensorFlow 1 as the deep learning framework. The initial learning rate is set to 0.001, and the weights are optimized using the Adam optimizer. A batch size of 16 and 100 epochs are used for training. The accuracy results during iterative training and validation are shown in Figure 4, where the light blue solid thin line, orange solid thick line, and black dotted line represent the training set accuracy, smoothed training set accuracy, and validation set accuracy, respectively. As observed in Figure 4, the accuracy exhibits a rapid increasing trend within approximately 10 epochs, showing significant improvement. After 40 epochs, the accuracy curve stabilizes, ultimately achieving 99.69% recognition accuracy for vortex beams with different $\ell$-values and $z$-values upon completion of 100 epochs. These results demonstrate the model's strong predictive capability. The non-smooth nature of the accuracy curve can be attributed to variations in turbulent phase updates during the training process.

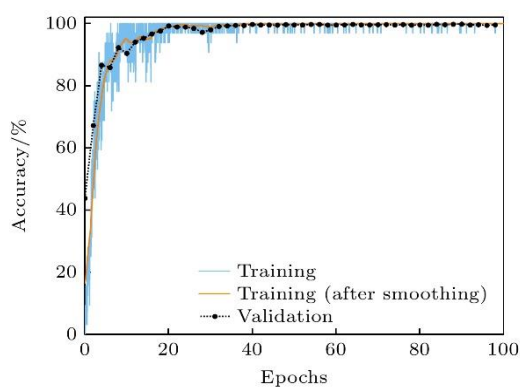

**Figure 4.** Accuracy of our trained deep learning algorithm.

The images in the test set are recognized by the trained model, and the confusion matrix is shown in Fig. 5. The vertical axis represents the input OAM topological charge $\ell$ and distance $z$, and the horizontal axis represents the machine-identified $\ell$ and $z$ values. Correct identifications, where the predicted OAM topological charge and propagation distance match the true values of the intensity distribution, appear along the diagonal positions. Off-diagonal elements with values greater than zero indicate deviations in

OAM topological charge recognition. Fig. 5(a) presents the normalized confusion matrix between the predicted propagation distance and the true propagation distance when $\ell=3.5$. Fig. 5(b) shows the normalized confusion matrix between the $\ell$ predicted value and the $\ell$ true value at $z = 75$ cm. It can be seen that almost all the tested OAM modes are correctly identified, and only one wrong prediction is located in the adjacent OAM state, which shows that small differences can also be clearly identified in this experiment. Furthermore, the method exhibits excellent generalization ability and robustness.

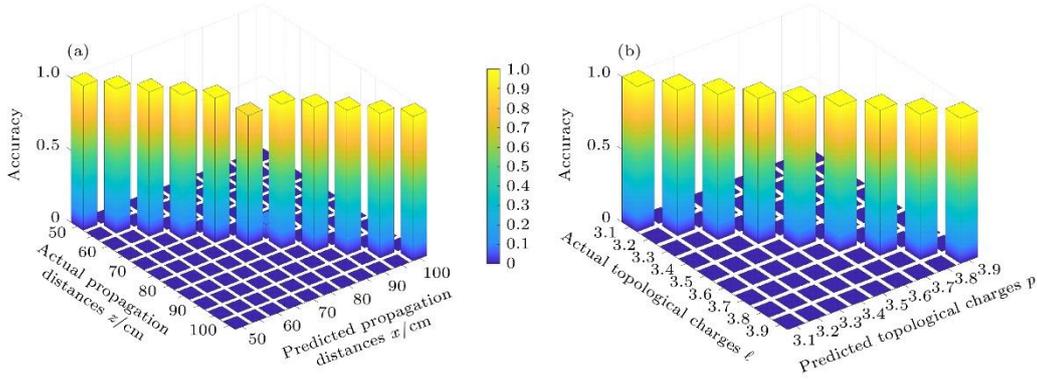

**Figure 5.** The confusion matrix of our trained deep learning algorithm: (a) The normalized confusion matrix between the predicted propagation distance and the true propagation distance for $\ell=3.5$; (b) normalized confusion matrix between predicted $\ell$ values and true $\ell$ values for z = 75 cm.

## 4. Discussion

This work is only a preliminary exploration of FOAM mode identification in turbulent environment at different propagation distances. The simulated turbulence belongs to the range of weak turbulence, and the accurate identification of FOAM modes and propagation distances under strong turbulence requires further investigation. Methods such as turbulence compensation or image reconstruction can be considered to improve the effect of turbulence.

In this paper, the range of distance measurement is 50-100 cm, which is relatively short. At this range, the influence of atmospheric turbulence on beam propagation is limited, differing significantly from the actual communication situation. However, the experimental results in this paper demonstrate the feasibility of our method. In the following work, we will study the machine learning recognition of the fractional order vortex light diffraction process at a long distance.

In the future, this work is expected to be applied in a variety of scenarios. Firstly, in the aspect of ranging, the diffraction characteristics of fractional vortex beams can be used to measure the propagation distance of beams. By analyzing the topological charge and diffraction distance of vortex beams, the position of objects can be accurately measured. Secondly, in free-space optical communication, the topological charge of fractional-order vortex beams can achieve higher-order mode multiplexing, increasing communication capacity; while the diffraction detection at the communication receiver end can enhance the stability of the communication system. Finally, in the application of optical tweezers, fractional vortex optical tweezers can manipulate the spatial position of particles, and by considering the diffraction effect, more precise control of particles using vortex beams can be achieved.

**5. Conclusion**

In this paper, a deep learning method based on ResNet is proposed and designed. By training the ResNet model to learn the mapping relationship between FOAM modes and diffraction intensity profiles, the FOAM modes can be accurately recognized at different propagation distances. This method takes into account the influence of atmospheric turbulence and other factors in the process of space transmission, so the recognition scheme can achieve high-precision recognition in special environments, and has the ability to distinguish ultra-fine FOAM modes and propagation distances that traditional methods can not achieve. The model provides a feasible method for high-precision recognition of FOAM modes with strong immunity, and can realize the recognition of propagation distance. It can recognize FOAM modes with a propagation distance interval of 5 cm and a mode spacing of 0.1, with an accuracy of 99.69%. This will help promote the practical application of FOAM mode in ranging, optical communication, micro-particle manipulation and other fields.

We would like to thank Dr. Chenglong You from Louisiana State University for helpful discussion.

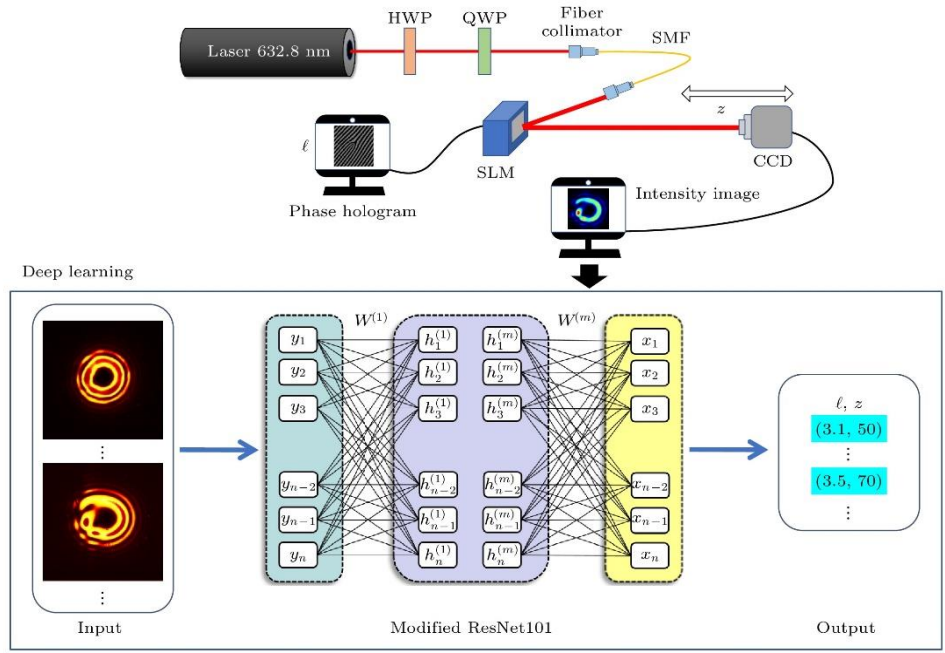

## References

[1] Shen Y, Wang X, Xie Z, Min C, Fu X, Liu Q, Gong M, Yuan X 2019 Light Sci. Appl. 8 90
[2] Bai Y, Lü H, Fu X, Yang Y 2022 Chin. Opt. Lett. 20 012601
[3] Zhang H, Zeng J, Lu X, Wang Z, Zhao C, Cai Y 2022 Nanophotonics 11 241
[4] Chen X, Wang S, You C, Magaña-Loaiza O S, Jin R B 2022 Phys. Rev. A 106 033521
[5] Guo Z, Chang Z, Meng J, An M, Jia J, Zhao Z, Wang X, Zhang P 2022 Appl. Opt. 61 5269
[6] Nye J F, Berry M V 1974 Proc. R. Soc. London, Ser. A 336 165
[7] Brygndahl O 1973 J. Opt. Soc. Am. 63 1098
[8] Allen L, Beijersbergen M W, Spreeuw R J C, Woerdman J P 1992 Phys. Rev. A 45 8185
[9] Senthilkumaran P, Sato S, Masajada J 2012 Int. J. Opt. 2012 1
[10] Wang J, Yang J Y, Fazal I M, et al. 2012 Nat. Photonics 6 488
[11] Kotlyar V V, Kovalev A A, Nalimov A G, Porfirev A P 2020 Phys. Rev. A 102 023516
[12] Zhu L, Tang M, Li H, Tai Y, Li X 2021 Nanophotonics 10 2487
[13] Nicolas A, Veissier L, Giner L, Giacobino E, Maxein D, Laurat J 2014 Nat. Photonics 8 234
[14] Otte E, Rosales-Guzmán C, Ndagano B, Denz C, Forbes A 2018 Light Sci. Appl. 7 18009
[15] Bu X, Zhang Z, Chen L, Liang X, Tang H, Wang X 2018 IEEE Antennas Wirel. Propag. Lett. 17 764
[16] Li X, Tai Y, Lü F, Nie Z 2015 Opt. Commun. 334 235
[17] Leach J, Courtial J, Skeldon K, Barnett S M, Franke-Arnold S, Padgett M J 2004

Phys. Rev. Lett. 92 013601
[18] Beijersbergen M W, Allen L, Van der Veen H, Woerdman J 1993 Opt. Commun. 96 123
[19] Zhou J, Zhang W, Chen L 2016 Appl. Phys. Lett. 108 111108
[20] Krenn M, Fickler R, Fink M, Handsteiner J, Malik M, Scheidl T, Ursin R, Zeilinger A 2014 New J. Phys. 16 113028
[21] Doster T, Watnik A T 2017 Appl. Opt. 56 3386
[22] Liu Z, Yan S, Liu H, Chen X 2019 Phys. Rev. Lett. 123 183902
[23] Jing G, Chen L, Wang P, Xiong W, Huang Z, Liu J, Chen Y, Li Y, Fan D, Chen S 2021 Results Phys. 28 104619
[24] Guo H, Qiu X, Chen L 2022 Phys. Rev. Appl. 17 054019
[25] Gao H, Zhang Z, Yang Y 2023 Appl. Opt. 62 5707
[26] Wu Y, Wang A, Zhu L 2023 Opt. Express 31 36078
[27] Zhao Y, Zhong X, Ren G, He S, Wu Z 2017 Opt. Commun. 387 432
[28] Zhou Z Y, Zhu Z H, Shi B S 2023 Quantum Eng. 2023 4589181
[29] Collins S A 1970 J. Opt. Soc. Am. 60 1168
[30] Bos J P, Roggemann M C, Gudimetla V S R 2015 Appl. Opt. 54 2039
[31] Glindemann A, Lane R, Dainty J 1993 J. Mod. Opt. 40 2381

[32] Bhusal N, Lohani S, You C, Hong M, Fabre J, Zhao P, Knutson E M, Glasser R T, Magaña-Loaiza O S 2021 Adv. Quantum Technol. 4 2000103
[33] Lü H, Guo Y, Yang Z X, Ding C, Cai W H, You C, Jin R B 2022 Front. Phys. 10 843932
[34] Fernando B, Habrard A, Sebban M, Tuytelaars T 2014 arXiv: 1409.5241 [cs.CV]
[35] Krizhevsky A, Sutskever I, Hinton G E 2017 Commun. ACM 60 84
[36] Zhang Z, Sabuncu M 2018 Advances in Neural Information Processing Systems 31 8778
[37] Kingma D P, Ba J 2014 arXiv: 1412.6980 [cs.LG]